\documentclass{emulateapj}

\shorttitle{HST observations of  PSR\,B1055$-$52}
\shortauthors{Mignani, Pavlov, \& Kargaltsev}

\newcommand{\hstn}{{\sl Hubble Space Telescope}}
\newcommand{\hst}{{\sl HST}}
\newcommand{\vltn}{{ Very Large Telescope}}
\newcommand{\vlt}{{ VLT}}
\newcommand{\fors}{{ FORS1}}
\newcommand{\forsn}{{ FOcal Reducer/low dispersion Spectrograph}}
\newcommand{\wfpc}{{ WFPC2}}

\newcommand{\gsc}{{ GSC2.3}}

\newcommand{\chan}{{\sl Chandra}}
\newcommand{\xmm}{{\sl XMM-Newton}}
\newcommand{\rosat}{{\sl ROSAT}}
\newcommand{\ein}{{\sl Einstein}}

\newcommand{\psr}{PSR\,B1055$-$52}

\begin{document}

\title{
Optical-UV spectrum and proper motion of the middle-aged pulsar B1055$-$52\thanks{Based on observations with the NASA/ESA \hstn, obtained at the Space Telescope Science Institute, which is operated by
AURA, Inc.\ under contract No NAS 5-26555.  Based on observations collected at the European Southern Observatory (ESO), La Silla and Paranal, Chile under programme ID  68.D-0407(A). }}

\author{R.\ P.\ Mignani\altaffilmark{1},
G.\ G.\ Pavlov\altaffilmark{2},
O.\ Kargaltsev\altaffilmark{3}
}
\altaffiltext{1}{ Mullard Space Science Laboratory, University College London, Holmbury St. Mary, Dorking, Surrey, RH5 6NT, UK; rm2@mssl.ucl.ac.uk}

\altaffiltext{2}{Department of Astronomy and Astrophysics, Pennsylvania State University, PA 16802, USA; pavlov@astro.psu.edu}

\altaffiltext{3}{Department of Astronomy, University of Florida, FL 32611, USA;
oyk100@astro.ufl.edu}

\begin{abstract}
PSR\,B1055$-$52 is a middle-aged ($\tau = 535$ kyr) radio, X-ray, and $\gamma$-ray pulsar showing  X-ray thermal emission from the neutron star (NS) surface.  A candidate optical counterpart to \psr\ was proposed  by Mignani and coworkers  based on {\sl Hubble Space Telescope} ({\sl HST}) observations performed in 1996, in one spectral band only. We report on {\sl HST} observations of this field carried out in 2008, in four spectral bands. The astrometric and photometric analyses of these data confirm the identification of  the proposed  candidate as the pulsar's optical counterpart. Similarly to other middle-aged pulsars, its optical-UV spectrum  can be described  by the  sum of a power-law (PL$_{\rm O}$) component  ($F_{\nu} \propto \nu^{-\alpha_{\rm O}}$), presumably emitted from the pulsar magnetosphere, and a Rayleigh-Jeans (RJ) component emitted from the NS surface. The spectral index of the PL$_{\rm O}$ component, $\alpha_{\rm O} = 1.05 \pm 0.34$, is larger than for other pulsars with optical counterparts. The RJ component, with the brightness temperature  $T_{\rm O}= (0.66 \pm 0.10)\,d_{350}^2\,R_{\rm O,13}^{-2}$ MK (where $d_{350}$ and $R_{\rm O,13}$ are the distance to the pulsar in units of 350 pc  and the radius of the emitting area  in units of 13 km),  shows a factor of 4 excess with respect to the extrapolation of the X-ray thermal component into the UV-optical. This hints that the RJ component is emitted from a larger, colder area, and  suggests that the distance to the pulsar is smaller than previously thought. From the absolute  astrometry of the \hst\ images we measured the pulsar coordinates  with a position accuracy of  $0\farcs15$. From the comparison with previous 
observations we measured the pulsar proper motion,  $\mu = 42\pm 5$ mas yr$^{-1}$, 
which corresponds to a transverse velocity $V_{t}  = (70\pm 8)\,d_{350}$ km s$^{-1}$.
\end{abstract}


\section{Introduction}

The radio pulsar B1055$-$52 (also known as PSR\,J1057$-$5226)   was discovered by Vaughan \& Large (1972). The period, $P=197$ ms, and its derivative, $\dot{P}= 5.83 \times 10^{-15}$ s s$^{-1}$, correspond to the spin-down age   $\tau = P/2 \dot{P}= 535$ kyr,  rotational energy loss rate (spin-down power)  $\dot{E} = 3.0 \times 10^{34}$ erg s$^{-1}$,  and surface magnetic field $B = 
1.1\times 10^{12}$ G.   Neither proper motion nor parallax measurements  have been reported for this pulsar so far.  Based on the  pulsar's dispersion measure,  DM=30.1 pc cm$^{-3}$, the distance was estimated as $d=1.53$ kpc for the Galactic free electron density model  by Taylor \& Cordes (1993),  
while the NE2001 model by Cordes \& Lazio (2002) yields $d=0.73\pm 0.15$ kpc.

PSR\,B1055$-$52 was one of the few radio pulsars detected by the \ein\ X-ray observatory 
(Cheng   \&  Helfand 1983).  Observations with {\sl EXOSAT} showed a thermal nature of its soft
X-ray emission, presumably originating from the neutron star (NS) surface
% with a temperature of $\sim 0.7$ MK 
(Brinkmann \& \"Ogelman 1987), while \rosat\  observations  found that at least two spectral components were needed  to fit the spectrum (e.g., a blackbody [BB] and a power-law [PL]) and discovered X-ray  pulsations  at the radio period (\"Ogelman  \& Finley  1993).    The similarities in the ages and X-ray properties with two nearby middle-aged pulsars, PSR\,B0656+14 ($\tau=110$ kyr)  and Geminga ($\tau =340$ kyr),  for which \rosat\ also detected thermal X-ray emission,  earned these three pulsars the nickname of ``The Three Musketeers'' (Becker \& Tr\"umper 1997).

The X-ray spectrum and pulsations of PSR\,B1055--52 were also studied with  {\sl ASCA}  (Greiveldinger et al.\ 1996; Wang et al.\ 1998) and  {\sl BeppoSAX} (Mineo et al.\ 2002).  
Recent observation  with the more sensitive \chan\ (Pavlov et al.\ 2002) and \xmm\  (De  Luca et  al.\ 2005) have shown that  the phase-integrated X-ray spectrum, similar to those of the other two ``Musketeers'',    can be fitted with three spectral  components: a cold BB ($T_{\rm C}\approx 0.8$ MK),
a hot BB ($T_{\rm H}\approx 1.8$ MK), and a PL component (photon index  $\Gamma\approx 1.7$, 
luminosity $L_{\rm X}^{\rm nonth} \sim 8\times 10^{30} d_{750}^2$ erg s$^{-1}$  in the 0.5--10 keV band, where $d_{750}=d/(750\,{\rm pc})$),  presumably emitted from the pulsar magnetosphere. Phase-resolved spectroscopy of the {\sl Chandra} and  {\sl XMM-Newton} data suggests that the changing projected emitting area of the hot component is responsible for the bulk of the phase variations (Pavlov et al.\ 2002; de Luca et al.\ 2005).
 
\psr\  was one of the seven $\gamma$-ray pulsars detected by the EGRET instrument aboard  the 
{\sl Compton Gamma-ray Observatory (CGRO)}  (Fierro et al.\ 1993; Thompson et al.\ 1996). It has also been studied by the {\sl Fermi} LAT detector (Abdo et al.\ 2010). Its 0.1--100 GeV luminosity, $L_\gamma
\sim 2\times 10^{34} d_{750}^2$ erg s$^{-1}$ (assuming an approximately isotropic emission),  implies an extremely high $\gamma$-ray efficiency, $\eta_\gamma\equiv L_\gamma/\dot{E}\sim 0.6 d_{750}^2$.

To study and interpret the magnetospheric and thermal emission from \psr, the X-ray and $\gamma$-ray data should be supplemented by optical data. However, optical observations with ground-based telescopes (Manchester et al.\ 1978; Cheng \& Helfand 1983; Bignami et al.\ 1988) failed to detect \psr\  because of the presence  of a bright ($V  \approx 14$) F-type field star  $\approx 4\arcsec$ from the radio pulsar position (dubbed Star A by Manchester et al.\ 1978). Thanks  to the  sharp resolution and  high  sensitivity  of the Faint Object Camera (FOC) aboard the \hstn\ ({\sl  HST}),  Mignani et al.\ (1997; hereafter M+97)  were  able  to identify a  faint ($U \sim 25$) probable  counterpart at the  radio pulsar position. Since the observation was taken in only one filter, the shape of the optical spectrum
of the candidate counterpart remained unknown, and even its identification with the pulsar was not certain. To confirm the identification, measure the spectral shape (in particular, separate the magnetospheric and thermal components, similar to the two other Musketeers  -- see, e.g. Kargaltev \& Pavlov 2007; KP07 hereafter), and measure the optical pulsations, new {\sl HST} observations were obviously required.  Our proposal to reobserve \psr\ with {\sl HST} was accepted for Cycle 16 (program 11154). These observations are described in  Section 2,  while the results are presented in Section 3, discussed in Section 4,  and summarized in Section 5.

\section{Observations and data analysis}

As no timing capabilities were available in the {\sl HST} Cycle 16,
we could only image the target in several filters. To observe \psr\ in the far-UV (FUV), most useful for studying the thermal component, we employed the Solar Blind Channel (SBC) detector of the Advanced Camera for Surveys (ACS).   The SBC is a CsI  Multi-Anode Microchannel Array (MAMA) photon-counting detector, which  is operated  in ACCUM mode  producing time-integrated images.  The detector is  sensitive to radiation in the 1115--1700 \AA\ wavelength range, and it provides a nominal field of view (FOV) of $34\farcs6  \times 30\farcs1$,  with a pixel scale of $0\farcs034 \times 0\farcs030$.  To filter out the geocoronal lines, we used the  longpass filter F140LP    (pivot wavelength $\lambda=1528$ \AA;  equivalent gaussian FWHM $\Delta \lambda=297$ \AA), which cuts off radiation at $\lambda \lesssim 1350$ \AA.  The pulsar was observed on 2008 February 13 over two consecutive spacecraft orbits for a  total integration time of 5569 s.  The data were reduced  
and calibrated (including the linearity correction, dark subtraction, flat-fielding) through the ACS data reduction pipeline ({\tt CALACS}), using the  closest-in-time calibration frames.  Geometric distortion correction, cosmic-ray  filtering, and exposure stacking were applied during the pipeline processing  using the {\tt Multidrizzle} task.  This task also applied a re-sampling of  the drizzled image to an even pixel size of $0\farcs025$.

We also  observed \psr\ in three optical bands with the Wide Field and Planetary  Camera 2 (\wfpc)\footnote{We  originally proposed the observations in the optical bands for the higher throughput ACS Wide Field Channel (WFC), but we had to switch to the \wfpc\ since the WFC was turned off in January 2007 because of problems with the ACS electronics.}  on 2008 March 8.  To exploit the maximum spatial resolution, the pulsar was centred in the Planetary Camera (PC)  chip, which has a pixel scale of $0\farcs045$ and a FOV of $35\arcsec \times 35\arcsec$. To maximize the spectral coverage,   we observed through the F450W ($\lambda=4557$ \AA; $\Delta \lambda=951$ \AA),   F555W ($\lambda=5442$ \AA; $\Delta \lambda=1229$ \AA), and  F702W ($\lambda=6917$ \AA; $\Delta \lambda=1381$ \AA) filters.  Observations were carried out during four consecutive orbits 
for the total integration times of 3600 s in the F450W filter and 1800 s  in each of the F555W and F702W filters.  For each filter, the observations were split in shorter, dithered exposures of 600 s  to filter out cosmic ray hits and bad pixels.  The data were processed through the \wfpc\ {\tt CALWP2} reduction pipeline for bias, dark,  and flat-field correction and flux calibration.  For each filter, we then combined single exposures with the IRAF task {\tt drizzle} to produce co-added and cosmic-ray-free images. 

We also used the archival FOC observations by M+97. They were carried out on 1996 May 11 in the F342W filter  ($\lambda=3402$ \AA; $\Delta \lambda=519$ \AA) during three spacecraft orbits   with the exposure times of 2771 s for the first  orbit and 3062 s for the second  and third orbits, corresponding to the total integration time of 8895 s. The data were obtained with the
high-resolution F/96 relay, with the FOV of $7\arcsec \times 7\arcsec$ and pixel scale of 0\farcs014. The data were retrieved from the  \hst\ science data archive\footnote{See \url{http://archive.stsci.edu}.} and on-the-fly recalibrated through the {\tt CALFOC} reduction pipeline with the updated reference files. 

\section{Results}

\subsection{Detection of the candidate counterpart}

The most recent published radio coordinates of \psr\  (Newton et al.\ 1981),
\begin{equation}
\alpha =10^{\rm h}  57^{\rm m} 58\fs 84\pm 0\fs03, \quad 
\delta= -52^\circ 26\arcmin 56\farcs3 \pm 0\farcs3,
\end{equation}
are for the epoch  of 1978.13
(equinox J2000). 
These coordinates, without proper motion values, are   reported in the ATNF radio pulsar  catalogue\footnote{See \url{http://www.atnf.csiro.au/research/pulsar/psrcat/}.} (Manchester et al.\ 2005), and they were also used by Weltevrede et al.\ (2010)  to compute the \psr\ radio ephemeris  for the timing of the  {\sl Fermi} $\gamma$-ray observations.   Since pulsars are known to have high velocities, \psr\ could move a few arcseconds in the 30 year span between the epochs of the radio position and our \hst\ observations  (2008.12 and 2008.18). For instance,  for an average radio pulsar velocity of 400 km s$^{-1}$ (e.g., Hobbs et al.\ 2005), we would expect a proper motion of  $110\, d_{750}^{-1}$ mas~yr$^{-1}$ and a position shift of $3\farcs4\, d_{750}^{-1}$ in 30 years,  in an unknown direction. Such an uncertainty hampers a direct identification  based on the pulsar's  position only. Fortunately, the presence of Star A in the immediate vicinity of the pulsar  can be used for the pulsar identification.

The $10\arcsec \times 10\arcsec$  cutout of the SBC image of the \psr\ field (left panel of Figure 1) shows 
the only two objects  detected in the entire SBC FOV, separated by a distance of 4\farcs41. From the comparison of this image with that obtained by M+97 with  FOC,   it is natural to assume
 that the northwestern and southeastern objects are Star A and the candidate pulsar counterpart, respectively. 

Indeed, the nominal SBC coordinates of the  northwestern source are  $\alpha=10^{\rm h}  57^{\rm m} 58\fs790$ and $\delta = -52^\circ 26\arcmin 53\farcs04$. The most accurate coordinates of Star A,  $\alpha_A=10^{\rm h}57^{\rm m}58\fs728 \pm 0\fs004 $ and $\delta_A=-52^\circ 26\arcmin 52\farcs45 \pm 0\farcs06$, at the epoch of our {\sl HST} observation, can be derived from the Third US Naval Observatory CCD Astrograph Catalog (UCAC3; Zacharias et al.\ 2010), with account for the Star A's proper motion, $\mu_\alpha=-8.9\pm 1.3$ mas yr$^{-1}$  and $\mu_\delta=7.8\pm 2.2$ mas yr$^{-1}$, given in the same catalog  and independently verified by us (see Appendix). As the difference between 
these and the nominal SBC coordinates  (0\farcs57 and $-0\farcs59$ in  right ascension and declination, respectively) is within the error budget  of the {\sl HST} absolute astrometry, we conclude that the SBC northwestern source is indeed Star A.

Applying the boresight correction to the SBC image, such that the coordinates of the northwestern source coincide with the UCAC3 coordinates of Star A, we obtain the following coordinates of the southeastern source
\begin{equation}
\alpha=10^{\rm h}57^{\rm m}58\fs954, \quad \delta=-52^\circ 26\arcmin 56\farcs37,
\end{equation}
with a nominal position  uncertainty of about $0\farcs07$. 
This uncertainty is mostly determined by the uncertainty of the UCAC3 absolute astrometry, thanks to the small angular distance (4\farcs41) between Star A and \psr, and
the $\sim 0.1^{\circ}$ accuracy on the \hst\ roll angle with respect to the equatorial reference frame.

The derived position is consistent
 with 
the candidate pulsar counterpart position in the FOC observation
($\alpha =10^{\rm h}  57^{\rm m} 58\fs832$, 
$\delta = -52^\circ 26\arcmin 56\farcs28$),   
within the uncertainty of the 
FOC position ($\approx 1\farcs0$)\footnote{Before 2002,  the pair of guide stars used  to compute the astrometric solution in the \hst\ focal plane   were selected from the GSC1.0 (Lasker et al.\ 1990),  which had a mean positional error of about $1\farcs0$. After 2002,  the guide  stars are selected from the \gsc\ (Lasker et al.\ 2008), which has an  improved mean positional error of about $0\farcs3$.} and the uncertainty caused by an unknown proper motion since the epoch of the FOC observation (1996.36). This suggests that we indeed detected the \psr\ candidate counterpart of M+97 in  our SBC image, while the extremely blue color (obvious from the comparison of the FUV (SBC) image with the optical (FOC and WFPC2) images; see Figure 1 and also Section 3.2) virtually proves that this object is  indeed \psr.

Using the offsets between the pulsar  counterpart and Star A measured in the SBC image (2\farcs02 and $-3\farcs92$ in right ascension and declination,  respectively), we looked for the pulsar counterpart  at the corresponding position in the WFPC2 images.  We detected it in both the F702W image (seen in the circle labeled ``SBC'' in the right panel of Figure 1) and in the F555W image but not in the F450W one.

\subsection{Photometry}

We measured the magnitudes and fluxes of the \psr\  counterpart in the \wfpc, ACS/SBC, and FOC images through customized  aperture  photometry using  the IRAF package {\tt digiphot}. 
For the WFPC2 F555W and F702W bands, we used a small source aperture with the radius $r=3$ pixels  
(0\farcs135)  to maximize the signal-to-noise ratio ($S/N$). We sampled the background in a surrounding annulus with the inner radius $r_b=5$ pixels and radial width $\delta r=10$ pixels;
the distance of 2 pixels from the source aperture was chosen to exclude the brighter portion of the point spread function (PSF) wings. We then applied the aperture correction using the fractional encircled energy (FEE) coefficients  (Holtzman et al.\ 1995), and the correction to compensate for the time- and position-dependent  charge  transfer efficiency (CTE)  losses of the  \wfpc\ detector (Dolphin 2009). 
For estimating the $3\, \sigma$ flux upper limit in the F450W band, we used the same $r=3$ pixels source aperture  but sampled the background in a wider annulus,  $\delta r=30$ pixels, around the expected pulsar position. 

For the SBC data, we  chose a source  aperture of $r=6$ pixels (0\farcs2) and a background annulus with  $r_b=20$ pixels and  $\delta r= 10$  pixels.  The large radial distance of 14 pixels between the source aperture and the background annulus was chosen to exclude the count excess over the PSF wings, which is seen from the comparison of the count distributions around the pulsar and around Star A (see the insets in the left panel of Figure 1).  The $3\,\sigma$ excess (e.g., $114\pm 38$ counts in the 0\farcs3--0\farcs7 annulus) cannot be ascribed to different PSFs of the pulsar and Star A because the  
latter is only one magnitude brighter than the pulsar in the SBC F140LP band,  and it is separated from the pulsar by only 4\farcs41. From these data alone,  we cannot exclude the possibility that the excess is a compact pulsar wind nebula (PWN), but, as the excess is not seen in other bands, it can also be just
a small-scale enhancement of the detector background at the pulsar position. Since the brightness profile of Star A is closer to that of the model PSF,  we used it  as a reference to compute the aperture correction  by fitting its growth curve  (i.e., the number of source counts within an aperture as a function of its radius).   

We also measured anew the magnitude and flux of the counterpart in the FOC image.  Following the approach described by Pavlov et al.\ (1997),  we chose an optimal source aperture from the measured growth curves  for each of the three orbits.  We found that the  growth curves for the first and third orbits  were consistent with each other and compatible with the nominal values of FEE at $\approx 3400$ \AA, while the growth curve for the second orbit  showed a peculiar behavior (e.g., no  saturation up to $r\sim 20$ pixels, which means an unusually broad PSF).  Therefore, we excluded the second orbit data from the analysis and chose $r=8$ pixels (0\farcs11), $r_b=10$ pixels, and $\delta r= 10$ pixels. 

In all cases, we applied  the countrate-to-flux conversion  by computing the photometric zero points  in the STmag system using  the  image keywords   {\tt PHOTFLAM}   and  {\tt PHOTZPT}, derived by the \hst\ data reduction pipelines.
To calculate the extinction-corrected fluxes, we estimated the  interstellar reddening  toward the pulsar, $E(B-V)=0.07$, using the hydrogen column density  $N_{H}=2.7 \times 10^{20}$ cm$^{-2}$, derived by
De Luca et al.\ (2005) from the X-ray fits, and using the correlation between  $N_H$ and $E(B-V)$ found by Predehl \& Schmitt (1995).  We then calculated the extinction coefficients using the  optical extinction curves by Fitzpatrick (1999) for the \wfpc\ and FOC passbands and the UV extinction curves by Seaton (1979) for the SBC passband. The magnitudes and the measured, $F_\nu^{\rm obs}$, and   dereddened, $F_\nu^{\rm der}$,  spectral fluxes are presented  in the last two columns of Table \ref{mag}. 

We plotted the dereddened spectral fluxes $F_\nu^{\rm der}$ versus frequency $\nu$ in Figure 2.  The shape of the broadband spectrum and, in particular,  the relatively high FUV brightness of the object  are inconsistent with it being  a non-collapsed star in the Galaxy or an extragalactic object. Therefore, the photometry proves unequivocally that the source is the optical counterpart of \psr.

It is clear from Figure 2 that a  simple one-component spectral model cannot fit the optical-UV spectral energy distribution (SED). However, similar to the optical-UV spectra of other middle-aged pulsars (e.g., 
KP07), the SED can be fitted by a model consisting of an (optical) power-law (PL$_{\rm O}$) component
and a Rayleigh-Jeans (RJ) component:
\begin{equation}
F_\nu^{\rm der} = f_{\rm O}\left(\frac{\nu}{\nu_0}\right)^{-\alpha_O} +\, \frac{2\pi}{c^2}\,\frac{R_{\rm O}^2}{d^2}\, kT_{\rm O}\,\,\nu^2 ,
\end{equation}
where $R_{\rm O}$ is the radius of equivalent emitting sphere, and $T_{\rm O}$ is the brightness 
temperature (both as seen by a distant observer).  At $E(B-V)=0.07$, the fitting parameters
are  $\alpha_{\rm O}= 1.05 \pm 0.34$,  $f_{\rm O}=0.112\pm 0.003$ $\mu$Jy at $\nu_0=1\times 10^{15}$ Hz, and $R_{\rm O}^2 T_{\rm O} = (510\pm 81) d_{750}^2$ km$^2$ MK (or $T_{\rm O}= (3.02 \pm 0.48)
\, d_{750}^{2}\,  R_{\rm O,13}^{-2}$  MK,  where $R_{\rm O,13}$ is  the radius in units of 13 km).
The PL$_{\rm O}$ component, which can be interpreted as  magnetospheric radiation,  dominates in the optical ($\lambda \gtrsim 3000$ \AA). The energy flux of this component is  ${\cal F}_{\rm O}^{\rm nonth} 
\approx 1.3 \times 10^{-15}$ erg cm$^{-2}$ s$^{-1}$ in the  3000--9000 \AA\ band. The RJ component dominates in the FUV and  is likely emitted from the NS surface. We will discuss implications of these results in Section 4.1.

\subsection{Astrometry}

\subsubsection{Absolute position of the pulsar}

The pulsar counterpart  position given by Equation (2) was obtained  from the  SBC astromery using the Star A position in the UCAC3 catalog.  Of course, using just one reference star may introduce a 
significant bias in the absolute target position. Fortunately, the WFPC2  has a much larger FOV than the SBC, allowing astrometry with larger numbers of reference stars.

For the WFPC2 astrometry, we used  two catalogs,  \gsc\  (Lasker et al. 2008) and 2MASS (Skrutskie et al. 2006).   Although the astrometric accuracy of these catalogs is somewhat lower than that of UCAC3, they contain more stars per given area, which improves the statistics of the astrometric fits. Moreover, most UCAC3 stars identified in the WFPC2 FOV are saturated, which prevents their use for the astrometric calibration. We used the F702W image, in which  \psr\ was detected with a  higher $S/N$ than in the F555W image. We produced the WFPC2 mosaic image with the STSDAS task {\tt wmosaic},  which also  corrects for the   geometric distortion. Then, we measured the centroids of the reference stars (13 stars in \gsc\ and 16 stars in 2MASS) in the WFPC2 pixel  coordinates   through  gaussian fitting  with the Graphical  Astronomy and  Image  Analysis (GAIA) tool\footnote{See \url{star-www.dur.ac.uk/$\sim$pdraper/gaia/gaia.html}.}  and used the catalog sky coordinates of these stars to  compute the pixel-to-sky coordinate  transformation with the code  {\tt ASTROM}\footnote{See \url{http://star-www.rl.ac.uk/Software/software.htm}.}.  This yielded astrometric fits with the rms  values $\sigma_{r} =  0\farcs17$  for both the \gsc\ and 2MASS stars. To these we added in quadrature the  uncertainties 
 of the  registration of  the \wfpc\ image on the  catalog reference frames, $\sigma_{\rm tr}= 0\farcs14$ and 0\farcs08, for the \gsc\ and 2MASS, respectively. Following Lattanzi et al.\ (1997), these uncertainties were estimated as $\sigma_{\rm tr}= (3/N_{\rm cat})^{1/2} \sigma_{\rm cat}$,  where $3^{1/2}$ accounts for  3 free parameters in the  astrometric fit,   $\sigma_{\rm cat}$  is  the  mean positional error of  the  catalog coordinates  (0\farcs3 and 0\farcs2 for the \gsc\ and 2MASS, respectively),  and $N_{\rm cat}$ is  the   number of  catalog stars used to compute the astrometric solution. Accounting for the  0\farcs15 and 0\farcs015  uncertainties on the   link of  the \gsc\ and 2MASS to the International  Celestial Reference  Frame (ICRF),  the overall  uncertainties of the  astrometry ($1\,\sigma$ position errors) are $\delta_{r}=0\farcs26$ and 0\farcs19, respectively.
 
We  applied these astrometric solutions to compute the sky coordinates  of \psr\ from its pixel coordinates. As it has been done for the reference stars,  we computed the pixel coordinates through a gaussian fitting,  with an uncertainty of  $\le 0\farcs01$ for the pulsar centroid, negligible 
in comparison with the overall uncertainty of  our absolute astrometry calibration.   We then obtained the
\gsc-based pulsar coordinates
\begin{equation}
\alpha =10^{\rm h} 57^{\rm m} 58\fs 961, \quad 
\delta = -52^\circ 26\arcmin 56\farcs17,
\end{equation}
with the position error $\delta_{r}=0\farcs26$, and the 2MASS-based coordinates
\begin{equation}
\alpha =10^{\rm h} 57^{\rm m} 58\fs 967, \quad
\delta = -52^\circ 26\arcmin 56\farcs31,
\end{equation}
with $\delta_r=0\farcs19$.
These two positions are consistent with each other and with the UCAC3-based position derived by using Star A as the only reference star in the SBC astrometry (Equation 2).

Since the coordinates in Equations (4) and (5) have been computed using different  catalogs and different reference stars, they can be considered independent and hence  can be averaged. The calculation of the weighted means yields
\begin{equation}
\alpha =10^{\rm h} 57^{\rm m} 58\fs 965, \quad
\delta = -52^\circ 26\arcmin 56\farcs26,
\end{equation}
with $\delta_r=0\farcs15$. Thus, the WFPC2 astrometry, re-calibrated with the \gsc\ and 2MASS, has provided 
%several estimates for 
the absolute pulsar coordinates at the epoch of 2008.18, which is separated by about 30 years from the epoch of the only published radio position\footnote{After completing our astrometric analysis, we became aware that an unpublished radio timing position was recently obtained (R.\ N.\  Manchester, private communication), which we found consistent with our own results.}. 

\subsubsection{Proper motion of the pulsar}

We used the  optical coordinates of \psr\  to  measure its proper  motion from the comparison with the 
radio coordinates at the 1978.13 epoch (Equation 1).  We note that,  since the \gsc\ and 2MASS, as well as UCAC3, are linked to the ICRF,  there is no a systematic offset  between the optical and radio coordinates. Using Equation (6),  we infer the displacement  $\Delta\alpha\, \cos\delta = +1\farcs14 \pm  0\farcs29$ and $\Delta \delta = +0\farcs04 \pm 0\farcs32$, which corresponds to the proper motion 
\begin{equation}
\mu_{\alpha} = +38\pm 10\,\,{\rm mas\, yr}^{-1}, \quad \mu_{\delta}= +1\pm 11\,\, {\rm mas\, yr}^{-1}
\end{equation}
in right ascenscion and declination.
Using the UCAC3-based positions (Equation 2) results in virtually the  same proper motion 
($\mu_\alpha= 35\pm 9$ mas yr$^{-1}$,  $\mu_\delta=-2\pm 10$ mas yr$^{-1}$), whose errors are dominated by the errors of the radio position (Equation 1). 

The pulsar proper motion can be independently measured through  {\em relative} \hst\/ astrometry, 
by comparing its position  with respect to Star A in the SBC image (epoch 2008.12) with that in the 
FOC image (epoch 1996.36).  Of course, measuring a relative proper motion  with only  one reference star comes with caveats.  Firstly,   Star A is slightly saturated in the FOC image and part of its PSF is outside the FOC FOV (see Figure 1 of M+97),  which  lowers the precision of the star  centroid determination.  To be very conservative, we assumed an uncertainty of 5 pixels (0\farcs07).  
The pulsar centroids were measured with a much higher precision of  $\lesssim 0.1$ pixels in both the SBC and FOC images  ($\lesssim 0\farcs0025$ and $\lesssim 0\farcs0014$, respectively), and the residual geometric distortion
(about 0\farcs004 and 0\farcs007 rms for the SBC and FOC, respectively -- see
Apell\'aniz \& Cox 2008 and Nota et al.\ 1996, respectively) is also much
smaller than the FOC centroiding uncertainty for Star A.

Secondly, with only one reference star, we have to rely upon the nominal values of the angles between the detector axes and the directions of right ascension and declination, which may introduce an error in the direction of the proper motion.  However,  this error is very small in our case, thanks to  the accuracy of  the detector position angle and the small separation of Star A from the pulsar (see \S\,3.1).
Last but not least, the uncertainty of the proper motion of the reference star  should be small enough not to hamper the measurement of the  relative pulsar proper motion.  We carefully verified it for Star A (see Appendix). 

Using the value of the  pixel scale for the two detectors and the nominal orientations of the  two images with respect to the equatorial  reference frame,  we found the displacement of the pulsar with respect to Star A during the time span of 11.76 years between the two \hst\ observations: $\Delta \alpha \, \cos \delta = 0\farcs61 \pm  0\farcs14$ and  $\Delta \delta = -0\farcs19 \pm 0\farcs14$. The uncertainty of the 
displacement is  dominated by the large uncertainty of the Star A centroid  in the FOC image.  This displacement corresponds to the proper motion  $\mu_{\alpha} = 52\pm 6$ mas yr$^{-1}$ and $\mu_{\delta}  = -16 \pm 6$ mas yr$^{-1}$. Correcting these values for the proper motion of star A 
as given in the UCAC3 catalog (see \S\,3.1), we obtained 
\begin{equation}
\mu_{\alpha} = 43\pm 6\,\, {\rm mas\, yr}^{-1}, \quad
\mu_\delta = - 5 \pm 6\,\, {\rm mas\, yr}^{-1}.
\end{equation}
This proper motion  agrees with that measured from the comparison of the radio and optical absolute coordinates. The weighted mean of the proper motion values determined by the two methods
%\footnote{We note that the effect of the CCD distortions over the very small {\em FOC} field of view is negligible in the comparison of the absolute and relative proper motions.}  
is
\begin{equation}
\mu_\alpha = 42\pm 5\,\, {\rm mas\, yr}^{-1}, \quad
\mu_\delta = -3 \pm 5\,\, {\rm mas\, yr}^{-1},
\end{equation}
or
\begin{equation}
\mu = 42\pm 5\,\, {\rm mas\, yr}^{-1}, \quad
{\rm P.A.} = 94^\circ \pm 7^\circ.
\end{equation} 
for the total proper motion and position angle (counted east of north). 
The large proper motion provides additional evidence that the M+07 candidate
is indeed the pulsar counterpart.

\section{Discussion}

\subsection{Pulsar spectrum}

We have shown in Section 3.2 that the optical-UV SED for \psr\  can be described by a sum of PL$_{\rm O}$ (magnetospheric) and RJ (thermal) spectra (see Equation 3). It is interesting to compare these optical components with the corresponding spectral  components at higher energies, particularly in soft X-rays where both the thermal and magnetospheric components are seen. As we have mentioned in Section 1,  the pulsar's X-ray spectrum  can be described by a model that consists of cold BB (BB$_{\rm C}$), hot BB (BB$_{\rm H}$), and X-ray power-law (PL$_{\rm X}$) components. The BB$_{\rm C}$ and BB$_{\rm H}$ components are presumably emitted from the bulk of the NS surface and polar caps, respectively, while the PL$_{\rm X}$ component, $F_\nu = f_{\rm X} (E/E_0)^{-\alpha_{\rm X}}$, represents the magnetospheric X-ray emission.  The parameters of the three components, as inferred from the \xmm\ data by De Luca et al.\ (2005), are the following: $T_{\rm C} =0.79\pm 0.03$ MK, 
$R_{\rm C}=12.3^{+1.5}_{-0.7}\, d_{750}$ km; $T_{\rm H}= 1.79 \pm 0.06$ MK, $R_{\rm H}=(0.46\pm 0.06)\, d_{750}$ km;  and $\alpha_{\rm X}=0.7\pm 0.1$, $f_{\rm X}=1.3^{+0.2}_{-0.1} \times 10^{-31}$ erg cm$^{-2}$  s$^{-1}$ Hz$^{-1}$ at $E_0 = 1$ keV, respectively. These components and their extrapolations into the optical-UV domain  are shown in the left panel of Figure 3. 

We see from Figure 3 that  the extension of the best-fit PL$_{\rm X}$ component into the optical
 overshoots the optical fluxes by a factor  of $\sim$ 3, similar to most of other pulsars detected in the optical (e.g.,  KP07; Mignani et al.\ 2010). However, in contrast to the majority of those
pulsars, the PL$_{\rm O}$ component is apparently steeper than  the PL$_{\rm X}$ component
($\alpha_{\rm O} = 1.05\pm 0.34$ vs.\ $\alpha_{\rm X}=0.7\pm 0.1$), and perhaps steeper than
the PL$_{\rm O}$ components in other pulsars (e.g., $\alpha_{\rm O} = 0.41\pm 0.08$ and $0.46\pm 0.12$ in PSR\, B0656+14 and Geminga, respectively; see KP07). Moreover, the extrapolation of the best-fit PL$_{\rm O}$ component into the X-ray range does not intersect the PL$_{\rm X}$ spectrum, being at least a  factor of $\sim 10$ lower at $E\gtrsim 1$ keV. Such behavior is unusual (a notable exception is PSR\,  B0540--69; Mignani et al.\ 2010), and it might suggest that different populations of ultrarelativistic electrons are responsible for the optical and X-ray emission of \psr. However, taking into
account the large uncertainties of the spectral slopes, it seems more plausible that   $\alpha_{\rm O}$ is substantially smaller than its best-fit value (e.g., $\alpha_{\rm O} \lesssim 0.5$), in which case
the extrapolation of the PL$_{\rm O}$ component can, in fact, be  smoothly connected with the PL$_{\rm X}$ spectrum.

Interestingly, the slope $\alpha_\gamma = 0.06\pm 0.1$ of the {\sl Fermi} LAT  spectrum, fit by a PL with an exponential cutoff ($F_\nu \propto E^{-\alpha_\gamma} \exp(-E/E_{\rm cut})$; Abdo et al.\ 2010), is even flatter than the X-ray slope. On the other hand, as we see from the right panel of Figure 3, the optical and $\gamma$-ray points can be connected by a PL spectrum with the slope $\alpha_{\rm O \gamma}\approx 0.46$,  which, however, goes slightly above the PL tail of the X-ray spectrum.
We cannot rule out the possibility that  further observations (especially in the IR-optical), supplemented by a joint  multiwavelength analysis, will show the magnetospheric spectrum of \psr\   to be similar to those of other pulsars with optical, X-ray, and $\gamma$-ray counterparts, including the other two Musketeers.

Another property of \psr\ to compare with those of other pulsars detected in the optical is  the relationship between the optical and X-ray luminosities of the magnetospheric emission. For instance, for the wavelength range 4000--9000 \AA, the flux in the PL$_{\rm O}$ component, ${\cal F}_{\rm O}^{\rm nonth} \approx 0.94\times 10^{-15}$ erg cm$^{-2}$ s$^{-1}$, corresponds to the luminosity $L_{\rm O}^{\rm nonth} = 4\pi d^2 {\cal F}_{\rm O}^{\rm nonth} =  6.3\times 10^{28}\, d_{750}^2$ erg s$^{-1}$ and optical efficiency  $\eta_{\rm O}\equiv L_{\rm O}^{\rm nonth}/\dot{E}=2.1\times 10^{-6}\, d_{750}^2$.  At the adopted distance of 750 pc, both the luminosity and efficiency  of \psr\ are considerably higher than
those of PSR\, B0656+14 and Geminga, for which $L_{\rm O}^{\rm nonth} =1.6\times 10^{28} d_{290}^2$, $\eta_{\rm O} =4.2\times 10^{-7} d_{290}^2$, and $L_{\rm O}^{\rm nonth} =4.8\times 10^{27} d_{250}^2$,  $\eta_{\rm O}=1.5\times 10^{-7} d_{250}^2$,  respectively (KP07). This hints that the actual distance to \psr\ is smaller, which is supported by the analysis of the thermal emission below. On the other hand, the ratio of the optical (4000--9000 \AA) and X-ray (1--10 keV) nonthermal luminosities, $L_{\rm O}^{\rm nonth}/L_{\rm X}^{\rm nonth} = 0.9\times 10^{-2}$ (which does not depend on the distance), is within the  same range of $\sim 10^{-3}$--$10^{-2}$ as for all other pulsars with optical
counterparts (Zavlin \& Pavlov 2004; Zharikov et al. 2006).
 
Let us now compare the optical-UV and X-ray {\em thermal}  components of the \psr\ spectrum.
The BB$_{\rm H}$ component is not seen in the optical because of the small emitting area, while
the extrapolation of the BB$_{\rm C}$ into the optical-UV is a factor of  $\approx 4$ below the observed RJ component. Such an ``optical-UV excess'' of the RJ component  has been  seen in the nearby radio-quiet isolated NSs (RQINSs), such as  RX\,J1856.5--3754 and RX\,J0720.4--3125 (e.g., Kaplan 2008, and references therein), and in the millisecond pulsar J0437$-$4715 (Kargaltsev et al.\ 2004).
It, however, has not been observed in middle-aged pulsars.  In particular,  the extrapolated BB$_{\rm C}$ component lies slightly {\em above} the optical-UV RJ component in Geminga (Kargaltsev et al.\ 2005),  while these components virtually coincide with each other in PSR\, B0656+14 (KP07).  

The nature of the optical-UV excess is not fully understood; a popular hypothesis is that it is due to
a nonuniform distribution of the temperature over the NS surface, such that the  BB$_{\rm C}$ 
component originates from a smaller and hotter region than the optical-UV RJ component (e.g., Pavlov et al.\ 2002).  Such an assumption is crudely equivalent to adding a ``very cold'' thermal component, BB$_{\rm VC}$,  which dominates in the UV but makes very little  contribution in X-rays. 
In other words, the product $R_{\rm O}^2 T_{\rm O}$ in the second term of Equation (3) is approximately equal to
$R_{\rm C}^2 T_{\rm C} + R_{\rm VC}^2 T_{\rm VC}$. 
Taking into account that $R_{\rm C}^2 T_{\rm C}\approx 119\, d_{750}^2$ km$^2$ MK and $R_{\rm O}^2 T_{\rm O}\approx 510\, d_{750}^2$ km$^2$ MK, we obtain $R_{\rm VC}^2 T_{\rm VC}\approx 391\, d_{750}^2$ km$^2$ MK  (or
$T_{\rm VC} \approx 2.31\, d_{750}^2 R_{\rm VC,13}^{-2}$ MK). Unfortunately, we cannot measure $T_{\rm VC}$ independent of $d^2/R_{\rm VC}^2$ in the RJ regime. We can, however, estimate an upper limit on $T_{\rm VC}$ from the requirement that  the BB$_{\rm VC}$ spectral flux extrapolated to
soft X-ray energies becomes so small, in comparison with the BB$_{\rm C}$,
that it does not affect the X-ray fit. We found that it occurs at $T_{\rm VC} \lesssim 0.45$ MK (at this temperature the contribution of 
 the BB$_{\rm VC}$ flux is $\lesssim 10\%$ above 0.3 keV -- see Figure 3, left).

Such a limit on $T_{\rm VC}$ implies a considerably smaller distance to the pulsar than that estimated from the pulsar's dispersion measure. Indeed,  from the above estimate on $R_{\rm VC}^2 T_{\rm VC}$ we obtain $d \lesssim 25.4 (R_{\rm VC}/1\,{\rm km})\,{\rm pc} < 330 R_{\rm NS,13}$ pc, where we took into account that $R_{\rm VC}$ is smaller than the NS radius, $R_{\rm NS}=13 R_{\rm NS,13}$ km.
Since the NS radius (as seen from infinity) can hardly exceed 20 km, the distance is  $< 500$ pc, and it can even be substantially smaller than this upper limit.

We should note that such estimates are, of course, model dependent. For instance, the temperature distribution over the NS surface can be smooth (i.e., there are no three distinct regions as we implicitly assumed above). Moreover, the spectrum of the thermal emission from a NS surface region can differ from the BB, 
which may lead to an excess of the  actually emitted optical-UV
 thermal component over the extrapolation of the X-ray thermal component even if the surface temperature is uniform (e.g., in the case of a light-element, H or He, NS atmosphere; Pavlov et al.\ 1996). However, fitting the model light-element  atmosphere spectra to the observed X-ray spectrum always leads to a lower temperature and a larger value of $R/d$ (i.e., to a smaller distance at a given NS radius)
than those obtained from the BB fits (Pavlov et al.\ 1995).  Therefore, it seems hard to avoid the conclusion on a smaller distance\footnote{Note a similar discrepancy between the DM-based distance estimate (750 pc, in the model by Taylor \& Cordes 1993) and the distance determined from parallax measurements ($288^{+33}_{-27}$ pc;  Brisken et al.\ 2003) for PSR\, B0656+14. Interestingly, Anderson et al.\ (1993) estimated the distance to be $280^{+60}_{-50}$ pc using the NS atmosphere models by Shibanov et al.\ (1993) and {\sl ROSAT} data.},  although its value cannot be determined accurately from the data available. We believe 200 pc and 500 pc are reasonable lower and upper limits, and we will scale the distance to 350 pc below. 

The conclusion that the distance is smaller than previously thought has important implications. 
For instance, the radius of the BB$_{\rm C}$ emitting region, $R_{\rm C} = 5.7^{+0.7}_{-0.3}d_{350}$ km, becomes substantially smaller than a plausible NS radius. It means that $T_{\rm C}=0.79\pm 0.03$ MK is {\em not} the temperature of the entire NS surface, and it should {\em not} be used for comparisons with the NS cooling models. This removes the  problem of \psr\ being too hot and luminous ($L_{\rm C,bol}=4.2\times 10^{32} d_{750}^2$ erg s$^{-1}$) for its age, in comparison with other middle-aged NSs and the predictions of standard NS cooling models (e.g., Yakovlev \& Pethick 2004).
 It also means that \psr\ is not a ``very slowly cooling NS'' (as suggested by Kaminker et al.\ 2002), and there is no need to invoke an unusually low NS mass to explain its thermal emission (see Yakovlev \& Pethick 2004 for references). To infer the true thermal luminosity and average surface temperature,
and compare them with the predictions of the NS cooling models, the parallax-based distance to the pulsar should be measured, and the distribution of the temperature over the NS surface should be
determined from the phase-resolved spectral analysis of the pulsar's soft X-ray and FUV emission. 

Another implication of the smaller distance is that the magnetospheric luminosities and efficiencies in various spectral bands are lower than previously thought, being close to the luminosities and efficiencies of other pulsars. 
For instance, if $d\sim 350$ pc, its $\gamma$-ray efficiency, $\eta_\gamma \sim 0.13\, d_{350}^2$ in the 0.1--100 GeV band (Abdo et al.\ 2010), although still rather high, becomes not so different from the efficiencies of other pulsars detected with {\sl Fermi} LAT, and the optical efficiency, $\eta_{\rm O}=4.6\times 10^{-7}\,d_{350}^2$, becomes comparable with those of PSR\,B0656+14 and Geminga. 

\subsection{Pulsar astrometry and kinematics}

\psr\  is the third rotation-powered pulsar,  after Geminga (Caraveo et al.\ 1998) and  PSR\,B0540$-$69 (Mignani et al.\ 2010),  for which an absolute position has been  determined with a sub-arcsecond accuracy  using optical astrometry techniques.   Knowing the astrometric position will be useful for the 
precise timing analysis of the pulsar.  

The proper motion we measured for the pulsar's optical counterpart  (Equation 11) corresponds to a tangential velocity  $V_t = (70\pm 8)\,d_{350}$ km s$^{-1}$.  This value is well below the average 400 km s$^{-1}$ for radio pulsars (Hobbs  et al.\ 2005), but it is close to that of the Vela pulsar  ($V_t \sim  65$  km s$^{-1}$; Caraveo et al.\ 2001).  

The proper motion in  Galactic coordinates,
\begin{equation}
\mu_l = 39\pm5 \,\, {\rm mas\, yr}^{-1}, \quad
\mu_b = 15\pm5 \,\, {\rm mas\, yr}^{-1},
\end{equation}
shows that the pulsar, whose current coordinates are $l=285.984^\circ$ and $b=+6.649^\circ$,  is moving away from the Galactic plane. We attempted to use  the proper motion  to locate the pulsar's birth place, identify a  parent star cluster (OB association), and estimate the actual age of the pulsar that can differ from its spin-down age. However, to calculate the pulsar's trajectory back in time in the Galactic potential, one should know the present distance (which is poorly known),
the tangential velocity (proportional to the distance), and the radial velocity $V_r$ (which is unknown).
Therefore, one has to calculate a large set of 
pulsar trajectories for various values of $d$ and $V_r$. Using the code developed by Vande Putte et al.\ (2010), we calculated the trajectories on a grid of 21 distances (in the range of 200 to 1200 pc) and 21 radial velocities ($-500$ to $+500$ km s$^{-1}$). 
We also selected a sample of young ($<100$ Myr) clusters/associations (de Zeeuw et al.\ 1999; Dias et al.\ 2002)  in a $25^\circ$ radius circle around the present position of the pulsar, 
extrapolated back in time their motion 
using the same code, and looked for closest approaches of the pulsar and cluster trajectories.
The results, however, turned out to be rather 
ambiguous because different values of $d$ and (especially) $V_r$ resulted in quite different candidate parent associations and pulsar ages
(e.g., we found 10 candidates for the age range of 400--700 kyr), 
and the uncertainty of ${\bf V}_t$ propagated back in time resulted in 
very large uncertainties on the minimum separations.  These uncertainties becomes even larger if one also accounts for the errors on the cluster proper motions, distances, and radial velocities. 
To obtain a more certain solution, we have to wait for the measurement of the pulsar's parallax and more accurate measurements of the proper motion, together with observational constraints on the pulsar's radial velocity, which could come, e.g. from the modelling of a possible bow-shock PWN created in the ISM
by the supersonically moving pulsar (see, e.g. Pellizza et al.\ 2005).

\section{Summary}

Using the \hst\ observations, we have confirmed that the candidate proposed by M+97 is indeed the optical-UV counterpart of \psr,
the tenth rotation-powered pulsar with a secured optical identification (see Mignani 2010 for a recent review). From multi-band photometry we  found that, similar to the  middle-aged pulsars PSR\, B0656+14 and  Geminga, its spectrum can be described  by the combination of  a PL$_{\rm O}$ component, with the spectral index $\alpha_{\rm O} = 1.05 \pm 0.34$,  and a RJ component, with 
brightness temperature  $T_{\rm O}=  (0.66\pm 0.10)\, d_{350}^2 R_{\rm O, 13}^{-2}$ MK. The PL$_{\rm O}$ component is steeper than those of PSR\,B0656+14 and Geminga  and, possibly, of all the other rotation-powered pulsars with known optical counterparts. Moreover, it is possibly steeper than the  X-ray PL component,  which might suggest that different populations of
relativistic electrons are responsible for the X-ray and optical-UV  magnetospheric emission.
To verify this assumption, IR  observations of \psr\ would be particularly useful, supplemented by a joint spectral fit  of the IR-optical, X-ray, and $\gamma$-ray components, using  multiwavelength spectral models for magnetospheric pulsar emission.

The observed RJ component exceeds by a factor of $\approx$4 the extrapolation of the X-ray thermal component into the UV-optical, unlike PSR\,B0656+14 and  Geminga but similar to RQINSs. It indicates that the temperature distribution over the NS surface is nonuniform, and the X-ray thermal component comes from a relatively small hotter region, while the main contribution to the optical-UV RJ component comes from a larger and colder area. It implies that the distance to the pulsar is considerably smaller
than estimated from the pulsar's dispersion measure and the models for  Galactic electron distribution. It also means that the conclusions inferred from the comparison of the X-ray temperature and thermal 
luminosity of \psr\ with the NS cooling models require a revision. 
New \hst\ observations in the UV and near-IR, including phase-resolved spectroscopy, 
are required to separate the thermal and magnetospheric components and infer the temperature distribution,
while the measurement of the 
radio parallax is crucial to determine  the actual distance and the NS radius.   

By recalibrating the  astrometry of the \hst\ images   with the \gsc, 2MASS and UCAC3 catalogs,
 we 
have measured the \psr\ absolute coordinates with a  radial position accuracy of 0\farcs15 at the epoch of 2008.18, about 30 years after the measurement of the only published  pulsar's coordinates. 
From relative astrometry between the \hst\ images taken 12 years apart,  we obtained the first measurement of the pulsar's proper motion,
$\mu =  42 \pm 5$ mas yr$^{-1}$ with a position angle of   $94^{\circ} \pm 7^{\circ}$, which corresponds to the transverse velocity 
$V_t= (70\pm 8) d_{350}$ km s$^{-1}$.    Further \hst\ and radio observations will allow one to measure the proper motion with a higher accuracy, to determine the distance from the  parallax measurement, and find the true pulsar age and the parent cluster/association.

\acknowledgments
Support for Program number 11154 was provided by NASA through a grant from  the Space Telescope Science Institute, which is operated by the  Association of Universities for Research in Astronomy, Incorporated,  under NASA contract NAS5-26555. RPM thanks 
A.\ C.\ Jackson for her contribution in the reduction of the \vlt\ data. We thank R.\ N.\ Manchester  for communicating the radio timing position of \psr\ before publication, A.\ Dolphin for useful discussions of the corrections for CTE  losses in WFPC2, D.\ Vande Putte for making available to us his code for Galactic orbit simulations, and M.\ van Kerkwijk for his valuable advice on the search for the pulsar's birthplace. We also thank the anonymous referee for the very careful reading of the manuscript and the useful suggestions that helped us to improve the presentation of the results.  The work by GGP and OK was partially supported by NASA grant NNX09AC84G.  

{\it Facilities:} \facility{HST (ACS, WFPC2)},\facility{VLT (FORS1)}

\appendix

\section{Star A proper motion}

Since the magnitude of  the Star A proper motion is crucial for determining   the pulsar's absolute position and proper motion   (see Sections 3.1 and 3.3.1),  we verified the UCAC3 value   ($\mu_\alpha=-8.9\pm 1.3$ mas yr$^{-1}$;  $\mu_\delta=7.8\pm 2.2$ mas yr$^{-1}$) against other astrometric catalogs.   Unfortunately, star A  is too faint   ($V  \approx 14.6$)  to be in the  Hipparcos  catalog ($V<12.4$; Perryman et al.\ 1997), while it is strangely  not in the deeper  Tycho-2 ($V<15.2$; H{\o}g et al.\ 2000) and  UCAC2 ($R<16$; Zacharias et al.\ 2004). It is, however, in  the USNO-B1.0 
catalog ($B<22$; Monet et al.\ 2003), which reports a  proper motion of   $\mu_{\alpha} =+30 \pm 4$ mas yr$^{-1}$ and  $\mu_{\delta}=0 \pm 6$ mas yr$^{-1}$ (much larger than in UCAC3), and in the PPMXL catalog  (Roeser et al.\ 2010), which gives   $\mu_{\alpha} =-13.8 \pm 8.2$ mas yr$^{-1}$ and 
$\mu_{\delta}=+4.5 \pm 8.2$ mas yr$^{-1}$,
consistent  with the UCAC3 value (albeit less accurate). 
 
As an independent test, we measured the relative proper motion of Star A. 
We used archival
\vltn\ (VLT) images of the \psr\ field taken with \forsn\  (\fors)  on 2002 March 21, which yields a time baseline of about 6 years with respect  to the \wfpc\ observations.  The images were taken in  high resolution mode  (0\farcs1/pixel) but with a windowing of the CCD   ($1\farcm6  \times 1\farcm6$ FOV).  A sequence of eighteen 140 s dithered exposures
was obtained through the Bessel B  filter  ($\lambda = 4290$\AA; $\Delta \lambda= 880$ \AA).  We applied standard CCD reduction with the {\em MIDAS} package and performed  the astrometric calibration as described in Section 3.3  (position error $\delta_{r}=0\farcs3$). Unfortunately,  the image quality 
of the \fors\ data (seeing $\sim 1\farcs2$) does not allow one to detect the pulsar in the halo of Star A.

We then compared the relative position of Star A measured in the \fors\  (2002.21) and \wfpc\ (2008.18) images. We used the F450W \wfpc\ image,  where Star A is not saturated and the pivot wavelength is similar to that of the Bessel B filter. 
We registered the \wfpc\ image  onto the \fors\ image, aligned along  right ascension and declination to better than  $\sim 0.05^{\circ}$,  by computing the transformation between the pixel coordinates of 15 stars 
seen in both images. This yielded an rms of  0\farcs015 along each of the  two axes, larger than the object centroid uncertainty (0\farcs003).   We thus determined the offset of Star A to be  $\Delta \alpha\, \cos \delta = -2.2\ \pm 15$ mas and  $\Delta \delta= 24.6 \pm 15$ mas, which corresponds to the $3 \sigma$ upper limit on the proper motion 
of 
$\sim 15$ mas yr$^{-1}$, consistent with both the UCAC3 and PPMXL values.

\clearpage

%{\scriptsize
\begin{table*}
{\scriptsize{
\begin{center}
  \caption{\hst\ photometry of the \psr\ counterpart. }
\begin{tabular}{lcccccccccc} 
\tableline\tableline
 Filter    & $\log \nu$  & $t$ & 
$r$ & $C_r$\tablenotemark{a} & $m_r$ & $\Delta m_r$  & $\Delta m_{\rm CTE}$ & $m_{\rm obs}$ & $F_\nu^{\rm obs}$     & $F_\nu^{\rm der}$\\ 
      & [Hz]      &  s  &  pix    &  counts     &              &                          &                                   &                      & $\mu$Jy     & $\mu$Jy \\
\tableline
WFPC2~F450W     & 14.818 (0.045)  & 600 & 
3 & $<$13.7          & $>$25.61		     & $-$0.27  &  $-$0.37  & $>$24.97	            & $<0.26$                & $<$ 0.34 \\
WFPC2~F555W     & 14.741 (0.049)  & 600 &
3 & $27.3\pm 5.0$   & $25.90\pm 0.20$ & $-$0.24  & $-$0.23  & 25.43  & $0.241\pm 0.044$ & 0.295 \\
WFPC2~F702W     & 14.637 (0.043)  & 600 & 
3 & $30.3 \pm 3.5$   & $26.46 \pm 0.13$ & $-$0.26  & $-$0.13  & 26.08  & $0.214 \pm 0.024$ & 0.249 \\ 
SBC~F140LP & 15.293 (0.042)  & 5569 & 
6 & $525 \pm 24$  & $22.83 \pm 0.05$ & $-$0.19  & \ldots            & 22.64  & $0.248 \pm 0.011$ & 0.417 \\ 
FOC~F342W  & 14.945  (0.033)  & 2916 & 
8 & $242 \pm 13$ & $25.18 \pm 0.06$ & $-$0.18  & \ldots            & 25.00  & $0.140 \pm 0.008$ & 0.190 \\ 	 
\tableline        
\label{mag}
\end{tabular}
\tablecomments{Columns report the camera and the filter, the pivot frequency and the band width (in parentheses) in logarithmic  units  (as derived from the SYNPHOT tables), the integration time of the 
average-combined frames, the photometry aperture radius $r$ in detector pixels, 
the total number of counts ($C_r$) measured within the aperture $r$,  the corresponding magnitude in the STmag system ($m_r$),  the aperture correction ($\Delta m_r$), the CTE correction  ($\Delta m_{\rm CTE}$), 
the observed magnitude  $m_{\rm obs}$, and the observed ($F_\nu^{\rm obs}$)  and extinction-corrected ($F_\nu^{\rm der}$) spectral fluxes.}
\tablenotetext{a}{For the WFPC2 and FOC observations, counts have been measured on the average-combined frames. For the FOC, we used only the average of the first and second exposure (see \S\,3.2).}
\end{center}
}}
\end{table*}

\begin{figure*}
 \includegraphics[height=8cm]{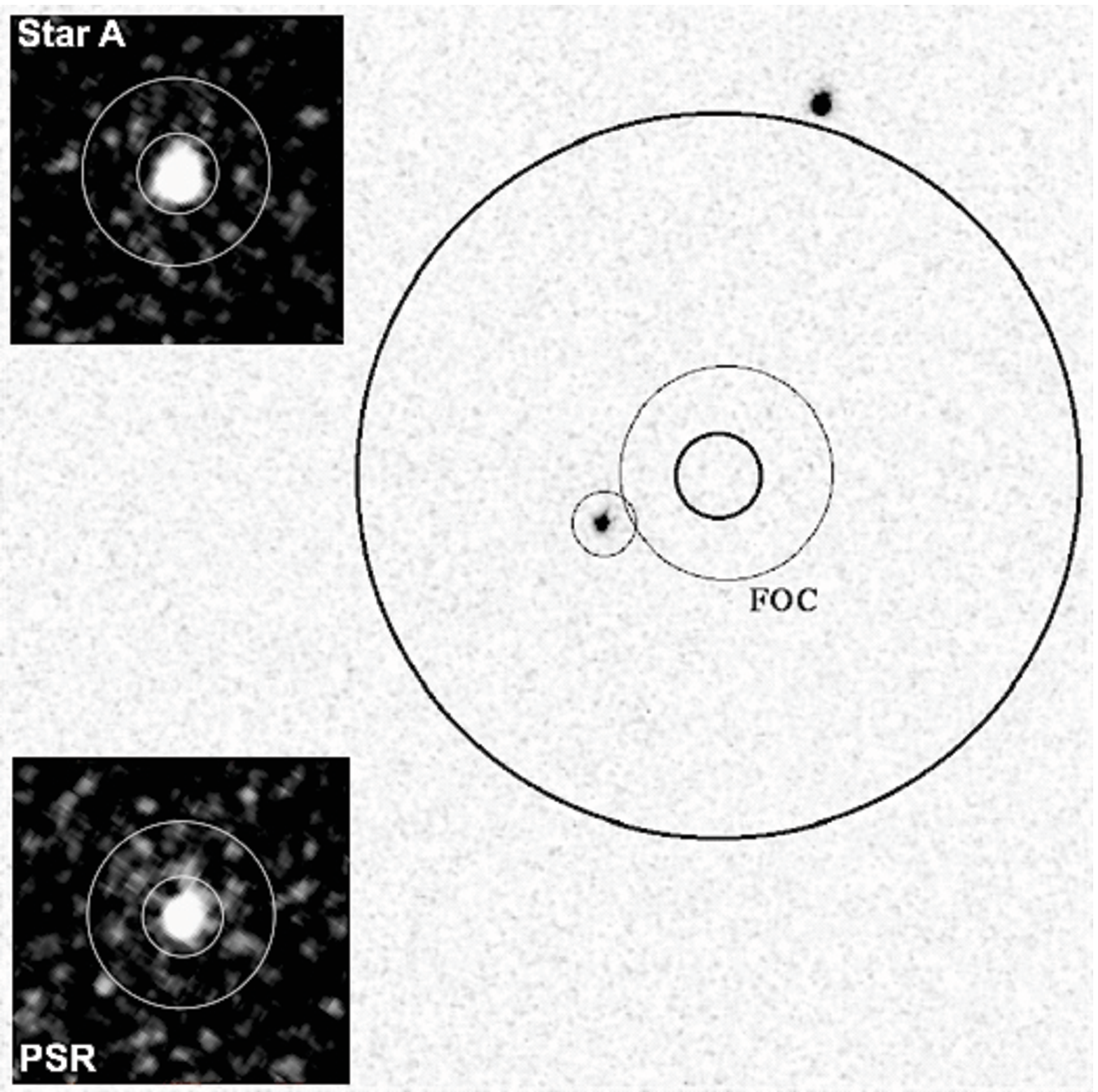}
   \includegraphics[height=8cm]{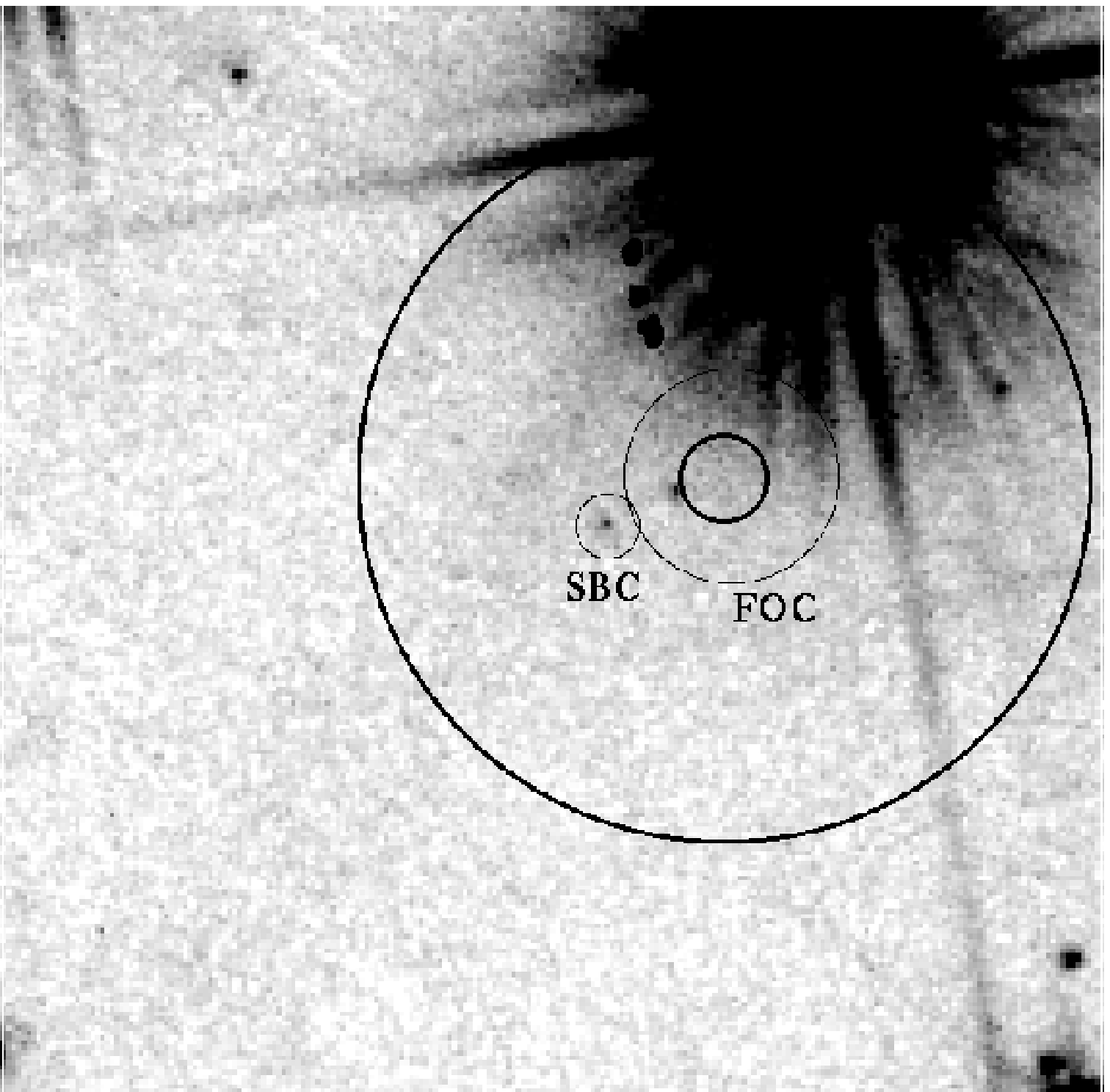}
  \caption{ SBC F140LP (FUV band; left panel)  and WFPC2 F702W  (red band; right panel) images  of the   PSR\,B1055$-$52 field   ($10\arcsec  \times 10\arcsec$). North is to the top, East to  the left. 
The two objects in the SBC image are the pulsar (in the smaller  thin circle  of 0\farcs3 radius, corresponding to a nominal uncertainty of SBC position)   and Star A (4\farcs4 northwest of the pulsar).   Zoomed images of the pulsar  and Star A, demonstrating the count  distributions in and around these 
sources, are shown in the insets; the radii of the circles in the insets  are 0\farcs3 and 0\farcs7.  The small thin circle in the WFPC2 panel, labeled  ``SBC'' is at the same position with respect to Star A as in the left (SBC)  panel. In each of the panels the smaller thick circle of 0\farcs45 radius is the  error circle around the radio pulsar position at the epoch of 1978.13  (Newton et al.\ 1981), while the larger thick circle of 3\farcs4 radius  encloses possible positions of the pulsar at the epoch of our \hst\  observations if the pulsar transverse velocity (in an unknown direction)   were 400 km s$^{-1}$, at $d=750$ pc.
The thin circle labelled ``FOC'' corresponds to the position of the \psr\  candidate counterpart of M+97; its radius of 1\farcs0 corresponds to the uncertainty of a nominal FOC position. See \S\,3.1 for more detail.
Notice the much higher relative brightness of the pulsar in the SBC image.  
}
\end{figure*}

\begin{figure*}
\begin{center}
\includegraphics[width=15cm]{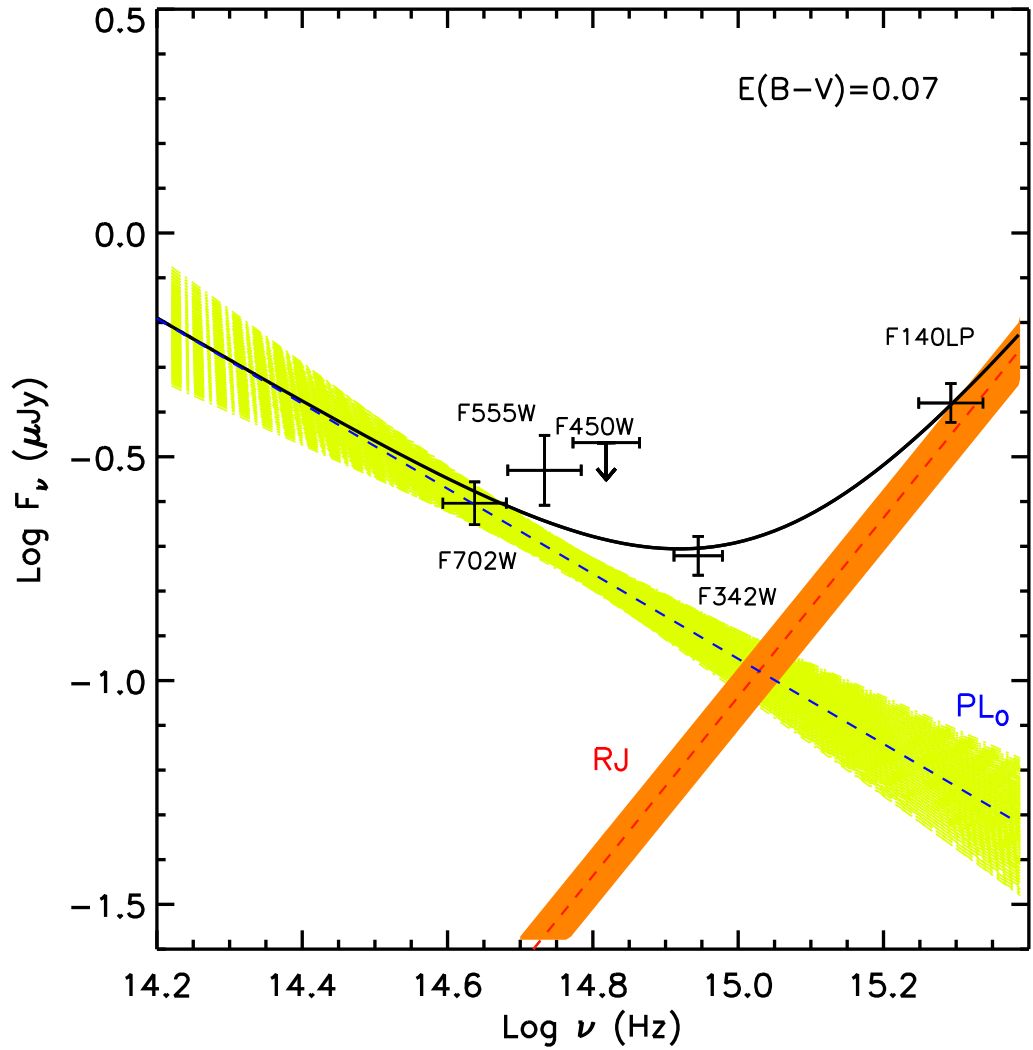}
\caption{Optical-UV spectrum of PSR\,B1055$-$52. The spectral fluxes have  been corrected for interstellar extinction assuming $E(B-V)=0.07$.
The dashed  lines correspond to the Rayleigh-Jeans (RJ; red) and  power-law  (PL$_{\rm O}$; blue) components of the best fit,  the solid line shows the sum of these components. The shaded areas show
$1 \sigma$ uncertainties of the fit.
}
\end{center}
\end{figure*}

\begin{figure*}
\hspace{-1cm}
\includegraphics[width=9cm]{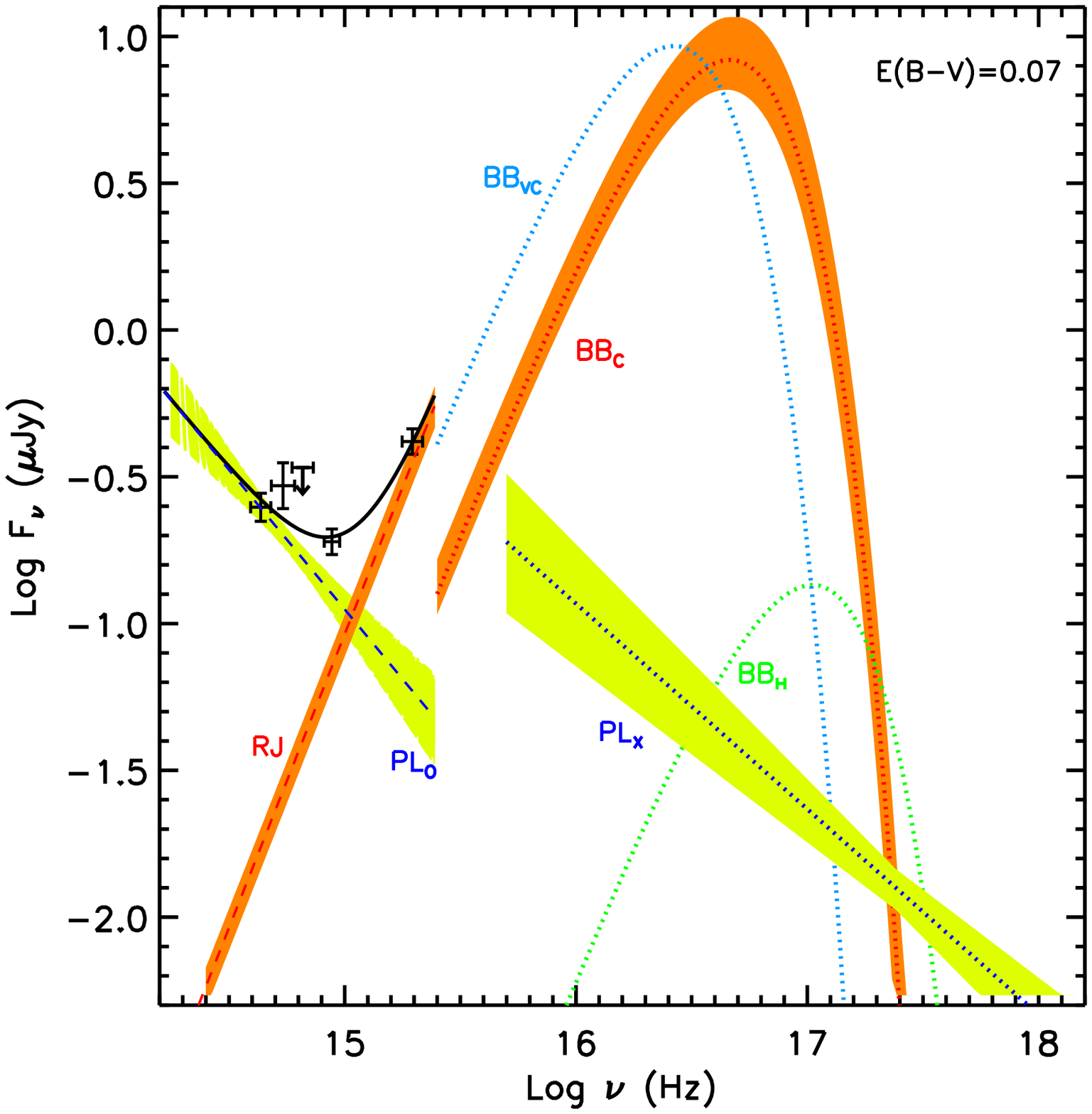}
\hspace{-0.5cm}
\includegraphics[width=9cm]{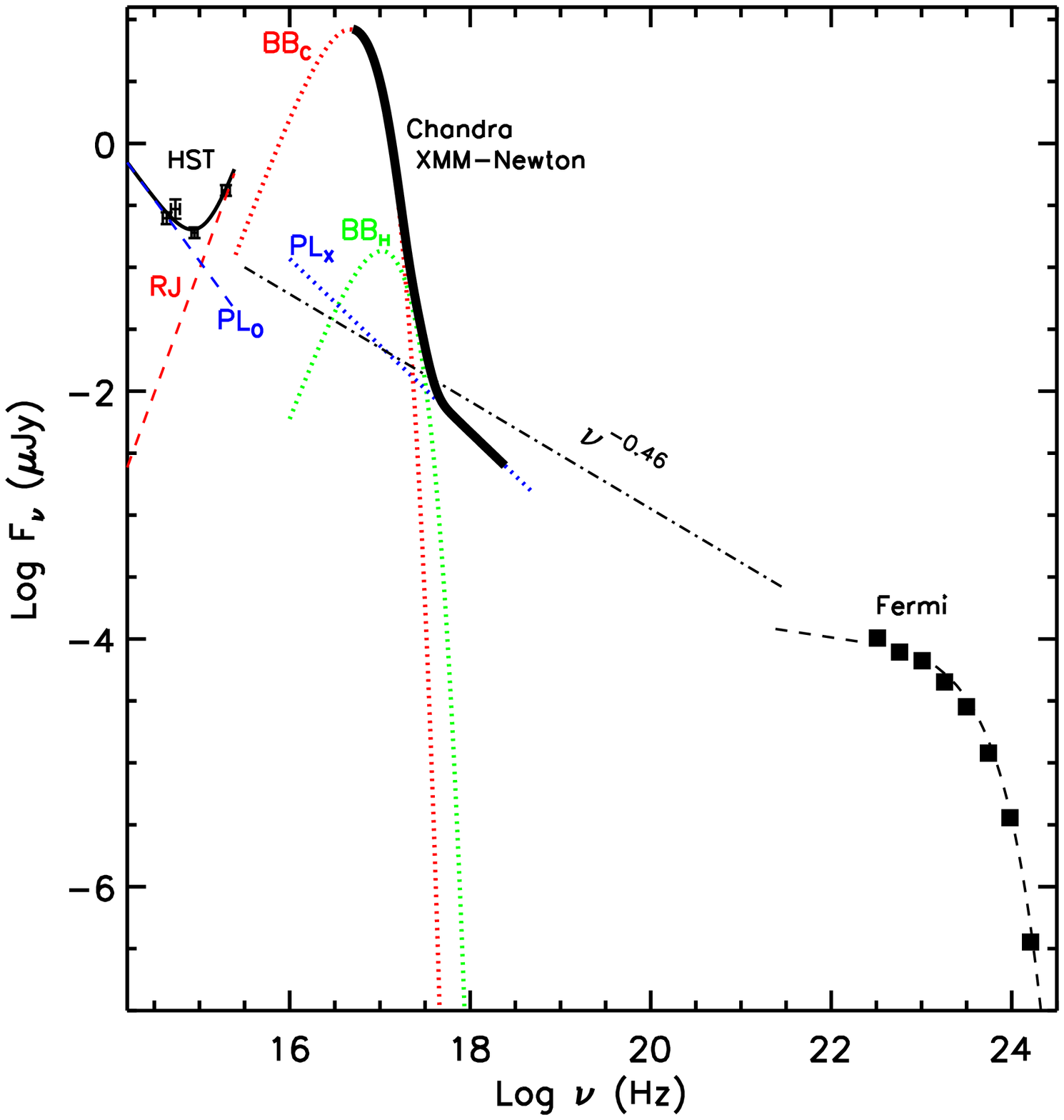}
\caption{{\em Left}: 
Comparison of the optical-UV and X-ray spectral models for PSR\,B1055$-$52. The X-ray fit for the BB$_{\rm C}$ + BB$_{\rm H}$ + PL$_{\rm X}$ model is taken from De Luca et al.\ (2005). The excess of the RJ component over the extrapolation of the X-ray BB$_{\rm C}$ component into the optical-UV range
could be explained by the presence of a ``very cold'' thermal component, BB$_{\rm VC}$. The dotted blue curve shows this component for the maximum allowed temperature $T_{\rm VC}=0.45$ MK.
The shaded areas show $1 \sigma$ incertainties of the fits. {\em Right}: Multiwavelength SED for PSR\,B1055-52, from optical to $\gamma$-rays, with spectral models for separate ranges. The $\gamma$-ray points are from Guillemot (2009), the spectral model from Abdo et al.\ (2010).
 The line $\nu^{-0.46}$ shows a PL spectrum approximately connecting the optical and $\gamma$-ray bands. See Section 4.1 for more details.
}
%\end{center}
\end{figure*}

\end{document}